\documentclass[conference]{IEEEtran}
	\IEEEoverridecommandlockouts
     \usepackage{fancyhdr,lipsum}

    \fancypagestyle{mahmood}{%
   \fancyhf{} 
   
   \fancyhead[C]{As similarly seen in IEEE Black Sea Conference on Communication and Networking, June 2022, Sofia Bulgaria.}
    }%

    \makeatletter
    \let\ps@IEEEtitlepagestyle\ps@mahmood
    \makeatother
    
	\usepackage{cite}
	\usepackage{amsmath,amssymb,amsfonts}
	\usepackage[noend]{algorithmic}
	\usepackage{graphicx}
	\usepackage{textcomp}
	\usepackage{xcolor}
	\usepackage{verbatim}
	\usepackage{afterpage}
	\usepackage{ragged2e}
	\usepackage{comment}
	\usepackage{breqn}
	\usepackage{amsmath}
	
\usepackage{amsmath}
\usepackage[noend]{algorithmic}
\usepackage{enumerate}
\usepackage{subfigure}
\usepackage{booktabs}
\usepackage{dsfont}
\usepackage{epstopdf}

\usepackage[justification=centering]{caption}
\usepackage[margin=1in,footskip=0.25in]{geometry}
	\def\BibTeX{{\rm B\kern-.05em{\sc i\kern-.025em b}\kern-.08em
		T\kern-.1667em\lower.7ex\hbox{E}\kern-.125emX}}
	
\begin{document}

	\title{IoT Threat Detection Testbed \\ Using Generative Adversarial Networks}
	\author{\IEEEauthorblockN{Farooq Shaikh$^{1}$, Elias Bou-Harb$^{2}$, Aldin Vehabovic$^{1}$, Jorge Crichigno$^{3}$, Ayseg$\ddot{\mbox{u}}$l Yayimli${^4}$, Nasir Ghani${^1}$\\
	\textit{$^{1}$Univ. of South Florida, $^{2}$Univ. of Texas San Antonio, $^{3}$Univ. of South Carolina, $^{4}$Valparaiso University}}
	}
	\maketitle

	\begin{abstract}
   The \textit{Internet of Things} (IoT) paradigm provides persistent sensing and data collection capabilities and is becoming increasingly prevalent across many market sectors. However, most IoT devices emphasize usability and function over security, making them very vulnerable to malicious exploits. This concern is evidenced by the increased use of compromised IoT devices in large scale bot networks (botnets) to launch \textit{distributed denial of service} (DDoS) attacks against high value targets. Unsecured IoT systems can also provide entry points to private networks, allowing adversaries relatively easy access to valuable resources and services. Indeed, these evolving IoT threat vectors (ranging from brute force attacks to remote code execution exploits) are posing key challenges. Moreover, many traditional security mechanisms are not amenable for deployment on smaller resource-constrained IoT platforms. As a result, researchers have been developing a range of methods for IoT security, with many strategies using advanced \textit{machine learning} (ML) techniques. Along these lines, this paper presents a novel \textit{generative adversarial network} (GAN) solution to detect threats from malicious IoT devices both inside and outside a network. This model is trained using both benign IoT traffic and global darknet data and further evaluated in a testbed with real IoT devices and malware threats. 
	\end{abstract}
	
  	\section{Introduction}
	\label{section:introduction}
	\indent 
	Continued advances in sensing, computing, and networking technologies have resulted the novel \textit{Internet of Things} (IoT) paradigm.  This approach uses a multitude of smaller embedded devices (augmented with sensing capabilities) to collect and relay information about their surroundings and environment. In many cases, these devices also interact with each other. This sensing information is then transferred to large datacenter sites (in the cloud) for further processing and data analytics purposes, i.e., to improve situational awareness and decision making processes. Indeed, IoT paradigms are seeing widespread traction across many diverse market sectors with many billions of devices already deployed, e.g., in domains such as transport, utility, building/infrastructure, manufacturing, healthcare, home automation, etc.

	Although IoT-based solutions offer tremendous benefits in terms of productivity and efficiency, they also introduce a plethora of security challenges. Namely, most IoT system manufacturers have emphasized cost reduction and rapid time-to-market over security support. As a result, many designs use very basic Linux or Unix \textit{operating systems} (OS) and have very limited (computational, storage) resource capabilities. Hence IoT devices are highly vulnerable to exploitation by malicious actors. Moreover, it is generally difficult (infeasible) to constantly patch these devices given their sheer scale of deployment, complicated access, and the lack of vendor support (update mechanisms).
	
	In light of the above, IoT infrastructures clearly represent a very large and growing cyberthreat surface.  As a result, unsolicited IoT-related activity on the Internet continues to grow at a steady rate (with {\tt telnet} being the most commonly exploited service). More importantly, hackers have also developed advanced IoT-specific malware to essentially recruit and coordinate much larger groups of devices and launch large scale \textit{distributed denial of service} (DDoS) attacks \cite{netsoft2020}.  By far the most notable example here is the {\tt Mirai} botnet which attacked Dyn DNS servers and caused massive Internet outages in 2016 \footnote{https://dyn.com/blog/dyn-statement-on-10212016-ddos-attack/}. The financial and service sectors were also impacted by this IoT botnet. Overall, {\tt Mirai} remains a major cyberthreat today, and hackers continue to evolve new variants with increasingly complex code structures and attack strategies \cite{ddosiniot}.  In addition, further methods are also being developed to target boot loader and firmware on IoT platforms. For example, the {\tt UbootKit} malware can manipulate the boot loader to grant root privileges to an attacker \cite{ubootkit}. Indeed, many of these malware types are becoming increasingly difficult to detect using traditional intrusion detection mechanisms.

    Despite these vulnerabilities, IoT technologies and solutions continue to gain traction across a wide range of private, commercial, and governmental domains. Therefore it critical to proactively identify and mitigate IoT-based threats before they materialize.  In response, researchers have proposed a range of  \textit{machine learning} (ML) schemes to detect anomalous behaviors in IoT domains \cite{bruno},\cite{svelte},\cite{deeplearningdistributed},\cite{iotthreatanalysis},\cite{computeintell},\cite{novelauto}.  However for the most part, many of these studies do not consider prominent IoT malware families and/or utilize realistic attack datasets.  Hence there is a pressing need to leverage real-world empirical network data to counter IoT threats.
    
    In light of the above, this paper presents a novel anomaly detection solution to identify threats from malicious IoT devices from both inside and outside the network. Namely, the scheme utilizes a \textit{generative adversarial network} (GAN) approach to model traffic distributions--for both benign and anomalous data--to identify malicious behaviors. This particular \textit{neural network} (NN) based scheme is chosen as it is very effective in modeling latent representations of data and reconstructing samples from their underlying distribution. An experimental real-world testbed is also developed to evaluate the proposed solution in realistic settings using the {\tt Mirai} and {\tt Bashlite} malware families. In particular, GAN models are trained using benign IoT data traffic collected from the testbed as well as global darknet data extracted from a large external network telescope ({\tt CAIDA} repository).  The latter traffic has been shown to be an effective indicator of Internet-scale malicious IoT device activity \cite{farooq1}. 

    Overall, this effort presents some key contributions. Foremost, it details one of the first known solutions which applies the advantages of GANs to the IoT security domain. This study also utilizes real-world darknet data (passive measurements) to further validate the efficacy of the model. Finally, the work builds a live operational testbed (using the most popular classes of IoT devices targeted and recruited by large-scale botnets) and evaluates the proposed solution using several IoT malware families. This paper is organized as follows. First, Section \ref{section:related} reviews a number of related works in the area of anomaly detection and ML techniques for both IoT and non-IoT networks. Next, Section \ref{section:setup} details the experimental testbed setup along with the proposed attack methodology. Section \ref{section:gan} then discusses the GAN model for anomaly detection of IoT devices. Finally, Section \ref{section:results} details the key research findings from the testbed performance evaluation study followed by concluding remarks and discussions on future research in Section \ref{section:conclusion}.

 
    \section{Related Work}
	\label{section:related}
    Anomaly detection algorithms have been widely used across a range of application domains, e.g., such as fraud detection, medical imaging, network intrusion detection, etc. The overall objective here is to detect any outliers in the data that deviate from expected or normal behaviors. Accordingly, the authors in \cite{anomalydetectionsurvey} present a comprehensive survey of anomaly detection applications and highlight the generic nature of many of these algorithms. Now many anomaly detection schemes have been used within the computer security domain. For example, \cite{anoamlydetectionnetwork2} uses a sequential matching algorithm to perform similarity measurements and identify deviations from past user behaviors. However, this method is ineffective against scanning and exploitation techniques directed at IoT devices. Meanwhile, \cite{anomalydetectnetwork3} details a statistical anomaly detection scheme that uses a multi-level hierarchical $K$-map model for network anomaly detection. However, this work does not consider IoT devices and the unique vulnerabilities of their ecosystem. Furthermore, the authors in \cite{anomalydetectionwebbased} detail an intrusion detection system for web-based attacks that analyzes server log files and produces anomaly scores for web requests. However this scheme is only designed for web applications and cannot detect network or link-layer layer attacks. Various signature-based defense mechanisms have also been proposed, i.e., by comparing network or application behaviors against a stored database of known malicious signatures \cite{anomalydetectionsurvey}. However, these schemes require detailed a-priori knowledge of malicious attacks, as well as constant updates to the threat signature database. As such, these requirements can be difficult to meet if attackers constantly adapt their behaviors to avoid detection.

    Now within the context of this study, \cite{bruno} presents an extensive survey on the latest developments in intrusion detection systems for IoT systems/devices. For example, the authors in \cite{svelte} propose an IoT network intrusion detection system called SVELTE. Specifically, this system is intended for 6LoWPAN IoT networks running the \textit{Routing Protocol for Low-Power and Lossy Networks} (RPL). Sample evaluation is also done using the Contiki OS \cite{ultra}, using bit pattern matching of IoT payloads to generate normal profiles and flag anomalous deviations. However, this solution is only suitable for small sensor- or control-based devices whose packets do not exhibit much variation in terms of their transmitted data.

    Meanwhile, \textit{machine learning} (ML) and \textit{deep learning} (DL) methods have also seen much traction in recent years owing to advances in computational hardware systems and optimized software packages. Specifically, open-source libraries such as {\tt TensorFlow} and {\tt Keras} provide extensive capabilities to rapidly design/test complex ML models. As a result, researchers have developed a range of intelligent ML-based security algorithms to address the large data volumes generated by IoT devices. For example, \cite{iwcms} presents a network traffic classification model which is trained and evaluated on darknet data using several common supervised ML algorithms, including random forest, gradient boosting, and Ada Boost. Meanwhile, the authors in \cite{netmine} develop association rules to detect anomalies, although this method still requires human expertise to draw meaningful conclusions.  Others have also used DL methodologies to detect unsolicited IoT behaviors. For example, the scheme in \cite{deeplearningdistributed} trains and tests a DL model using the well-known NSL-KDD dataset. However, this dataset is not specific to IoT devices and hence is not necessarily reflective of their specialized attack traffic, as noted in \cite{analysiskdd}. Meanwhile, \cite{iotthreatanalysis} also uses simulated data to train a NN to detect malicious IoT devices. However, the focus here is on detecting DoS/DDoS attacks against IoT sensor nodes themselves, and the lack of realistic attack traffic is also problematic.  Indeed, the latter issue is a major concern when trying to develop real-world operational intrusion and anomaly detection systems.
    
    Additionally, the study in \cite{computeintell} uses a computational intelligence algorithm to generate behavioral profiles and flag deviations. However this effort only considers wireless systems. Meanwhile, the authors in \cite{nabiot} present a scheme to train autoencoders for every IoT device connected to a network. Hence when a device is infected, its previously-trained autoencoder can be used to identify the anomaly. Clearly, this approach has high overheads as separate autoencoders must be trained for every IoT device. Moreover, this solution does not account for external threats since the autoencoder can only identify anomalies for the device for which it was trained. Other efforts in \cite{novelauto} have also looked at using variational autoencoders for anomaly detection. Nevertheless, this work does not consider prominent malware families such as {\tt Mirai}, {\tt Reaper}, etc.

    In light of the above, the proposed ML-based solution leverages GANs for anomaly detection in IoT settings.  Namely, this particular type of NN is well-suited for scenarios with sparse training datasets. GANs can also excel at generating latent representations of high-dimensional complex datasets. Hence this algorithm is used here to model both \textit{normal} traffic behaviors for clean IoT devices as well as \textit{malicious} traffic behaviors for compromised/infected IoT devices (from darknet data).  The GAN models are then used to detect anomalous activities arising from infected devices or network scanning activities. Overall, the key objective here is to stop attacks in the earlier scanning stage, thereby minimizing potential damage.  Further details are now presented.
\begin{figure*}[h]
    \centering
    \includegraphics[width=5.25in, height=3.0in] {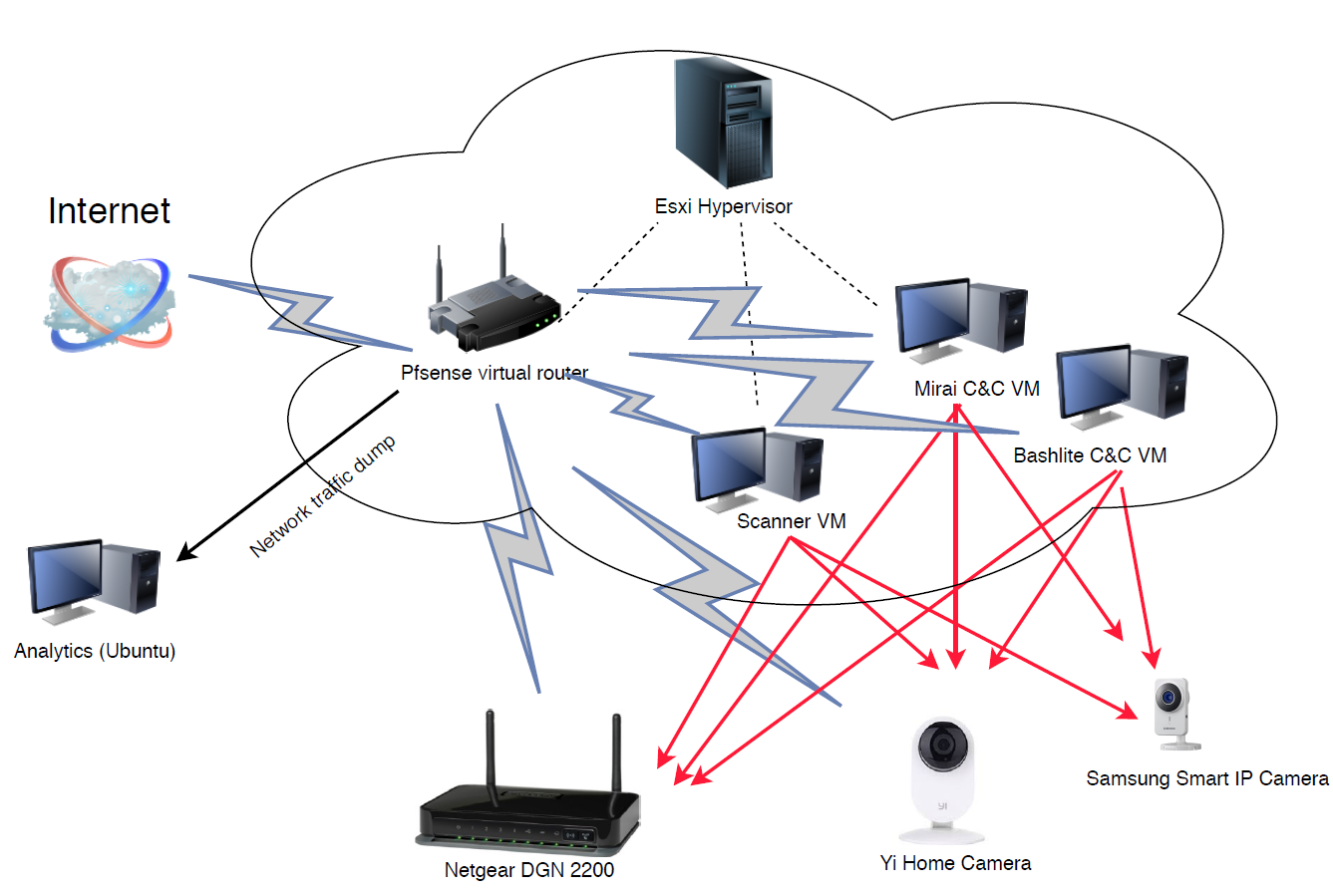}
    \caption{Overview of IoT testbed setup}
    \label{infrass}
\end{figure*}
\section{Experimental Testbed}
\label{section:setup}
    Before presenting the IoT anomaly detection solution, the experimental IoT testbed is detailed first, Figure \ref{infrass}. The overall objective here is to build a realistic evaluation environment comprised of physical IoT devices and then use it to study actual IoT-based malware attacks.  Accordingly, two major IoT malware types are considered, i.e., {\tt Mirai} and {\tt Bashlite}. These DDoS families are capable of generating massive attack traffic volumes and have caused serious Internet-scale outages \cite{ddosiniot}. Further details on the testbed systems and attack methodologies are now presented.
    
    \subsection{Testbed Systems}
    Overall, two main types of IoT systems are chosen to best represent the categories of devices commonly targeted by malicious actors, i.e., Wifi-based \textit{access point} (AP) routers and webcam \textit{digital video recorders} (DVR) systems. Foremost, Wifi-based routers are very common in many private and enterprise settings and represent lucrative attack targets. Additionally, DVR devices are also ubiquitous and have been regularly exploited by large botnets to launch massive Internet-scale DDoS attacks \cite{farooq1}. Many of these devices also offer a high level of security against physical threats, i.e., since they are usually located in secured indoor areas.

    In light of the above, three commercial IoT devices are incorporated into the testbed.  Foremost, the Netgear DGN 2200 Wifi router is chosen, as this specific model is known to be highly susceptible to remote code execution exploits \footnote{https://www.exploit-db.com/exploits/31617/}. For example, one these vulnerabilities includes the simple HTTP POST command (with malicious code) to gain administrative access to the device. Meanwhile, two types of webcams are also chosen, including the Yi 1080p Home Camera and a Samsung Smart IP camera. Again, such camera devices are very prevalent in many private and commercial settings and also use a basic Linux-based OS. Since both the {\tt Mirai} and {\tt Bashlite} botnets use brute forcing of default manufacturer credentials, the above platforms are well-suited for recreating such scenarios. Indeed, default usernames/passwords are readily available on the Internet for many existing IoT systems. This provides a relatively easy attack vector for hackers since most users do not change these default manufacturer settings. For example, the {\tt Mirai} source code contains a list of 50-60 default usernames/passwords to try to gain access to vulnerable devices. Once compromised, architecture-specific malware binaries are downloaded onto the IoT platforms to launch future DDoS or scanning attacks.

    Now many large IoT botnets are also managed by external \textit{command and control} (C\&C) servers, i.e., which initiate various commands and attack sequences on the infected devices. To effectively model these frameworks in the testbed, the {\tt VMware ESXi} hypervisor tool is used to setup two {\tt Ubuntu} \textit{virtual machines} (VM). The C\&C server code bases for the two different IoT malware families ({\tt Mirai} and {\tt Bashlite}) are then run on these VM instances, limiting the need for separate physical machines. Overall, the {\tt VMware ESXi} virtualization tool provides a simplified user interface for deploying VMs and can support most OS types. This tool also provides a seamless separation between the OS and underlying hardware, along with centralized management and reliable backup/restore capabilities. Note that the testbed also creates different subnets for the IoT devices and C\&C control servers in order to better emulate real-world networking setups.
    
    Many malware designs also use separate distribution servers to disseminate their malicious binaries. Accordingly, another VM is setup to run the {\tt Apache} webserver for such purposes.  Namely, the C\&C servers (running on the {\tt VMware ESXi} hypervisor) instruct the compromised IoT devices to contact this web server to download malware binaries specific to their architecture (using the WGET or FTP commands). Finally, the {\tt Metasploit} tool \footnote{https://metasploit.help.rapid7.com/docs} is also used to launch actual scanning and reconnaissance attacks against the testbed IoT devices. Specifically, a {\tt Kali Linux} VM is chosen for this purpose (scanner VM, Figure \ref{infrass}) since this Debian-based OS includes the {\tt Metasploit} module with many different payload, auxiliary, and post exploitation capabilities. Note that this tool has also been used by penetration testers to identify network or system weaknesses, see \cite{kali}. 
    
    Finally, it is important to incorporate network firewall capabilities into the testbed as these systems are widely deployed in real-world settings. Hence the {\tt pfsense} virtual router and firewall solution \cite{pfsense} is chosen to manage the testbed networking setup (instead of actual physical devices). This package offers increased flexibility and scalability, as well as centralized network management support. The {\tt pfsense} packet management system also interfaces with other tools such as {\tt Snort} (open-source intrusion detection system) and {\tt OpenVPN}. As such, this software provides a global view of all network communications and can be used to implement access control and other security policies via its simple web interface.  For example, the virtualized {\tt pfsense} router can be used in conjunction with {\tt Snort} to block any flagged traffic, thereby emulating firewall actions.

\subsection{Attack Methodology}
\label{attack}
    Malware developers are continually evolving new attack strategies to counteract defense mechanisms put in place after the emergence of {\tt Mirai} and other IoT botnets. Namely, advanced techniques are being used to infect devices \cite{iotevolvepaper1}, unlike earlier brute forcing mechanisms. Accordingly, a list of potential cyberattacks against the testbed IoT devices is shown in Table \ref{tsig1}.  Now in order to recreate such scenarios, the {\tt Kali Linux} VM is used to generate some attacks via its {\tt Nessus} module, e.g., such as checking for default credentials, web application scanning, etc. Since many IoT devices also have web management interfaces, {\tt Nessus} can also be used to run web application scanning and help attackers identify potential weaknesses in web interfaces. Hence this tool is used to launch about two thirds of the cyberattacks in Table \ref{tsig1}, including command injection and remote code execution. Furthermore, the {\tt Nmap} tool is also used for network scanning purposes, as it provides important information on the device OS, architecture, list of open ports, etc. Specifically, this tool can emulate the reconnaissance phase of a malware attack and provide hackers with critical information to improve their attack targeting.
    
    Now as noted earlier, some versions of the Netgear DGN 2200 router are known to be vulnerable to multiple exploits (and a brief description of some of these vulnerabilities is given in Table \ref{tsig}). However, detailed testing of the particular system purchased indicated that the latest firmware was no longer susceptible to these exploits. Hence in order to overcome this issue, the firmware was manually downgraded to a version that was vulnerable to the aforementioned attacks.
    
    Overall, the IoT testbed design is quite flexible and can be used to model a range of IoT cyberattacks, as summarized in Table \ref{tsig}.  For example, consider a \textit{man-in-the-middle} (MITM) attack on the (wireless) Yi 1080p Home Camera. Here an \textit{address resolution protocol} (ARP) cache poisoning scenario can be considered where an adversary poses as a legitimate Wifi \textit{access point} (AP) by advertising a valid gateway \textit{media access control} (MAC) address. Once the camera connects to this ``fake'' AP, the attacker can then decipher all communications from this device. Hence in order to recreate this MITM attack scenario, a wireless adapter can be used to sniff all wireless network traffic and perform packet injection. The {\tt Wifite} module (attack tool) in the {\tt Kali Linux} VM can then be used to emulate a malicious entity posing as a legitimate AP. 

\begin{table*}[!ht]
	\centering
	\caption{Cyberattacks against IoT testbed devices}
	\label{tsig1}
	\begin{tabular}{|l|c|c|c|}
		\hline \hline
		\textbf{Attack} &  \textbf{Netgear DGN2200} & \textbf{Yi 1080p Camera} & \textbf{Samsumg IP Camera} \\ [0.5ex] 
		\hline\hline
		 Nmap scanning & Yes & Yes & Yes \\ 
		\hline
		 Command injection & Yes & No & No\\
		 \hline
		 Nessus scanning & Yes & No  & Yes\\
		 \hline
		 Remote code execution & Yes & No & No \\
		 \hline
		 Mirai & Yes & Yes & Yes \\
		 \hline
		 Bashlite & Yes & Yes & Yes \\
		 \hline
		 MITM & No & Yes & No \\
		\hline \hline
	\end{tabular}
\end{table*}
\begin{table*}[ht]
	\centering
	\caption{Exploits against Netgear DGN 2200 Wifi router}
	\label{tsig}
	\begin{tabular}{|l|l|c|c|}
		\hline \hline
		\textbf{Vulnerability} & \textbf{Description}  \\ [0.5ex] 
		\hline\hline
		Command injection & Using special HTTP POST request \& login credentials \\ 
		\hline
		Remote command execution & Access to default user account   \\
		\hline
		Cross-site request forgery & Unauthenticated remote code execution \\
		\hline \hline
	\end{tabular}
\end{table*}
\begin{table*}[!h]
	\centering
	\caption{Feature extraction from raw packet data}
	\label{tsig3}
	\begin{tabular}{|l|l|c|}
		\hline \hline
	\textbf{Feature} &  \textbf{Description} \\ [0.5ex] 
		\hline\hline
		 Number of packets &  Mean \& std (total number of packets in both directions) \\ 
		\hline
		 Packet length & Mean \& std (length of the packets in both directions) \\
		 \hline
		 Number of unique ports & Ports used for communication in both directions  \\
		 \hline
		 Packet inter arrival time & Mean \& std (time between arrival of packets in both directions)\\
		 \hline
		 TCP  & 1 if using TCP, 0 if using UDP \\
		 \hline
		 PSH flag & Number of packets with PSH flag set \\
		 \hline
		 URG flag & Number of packets with URG flag set \\
		 \hline
		 Idle time &  Duration for which the connection was idle \\
		 \hline
		 Active & Duration for which the connection was active \\
		 \hline
		\hline \hline
	\end{tabular}
\end{table*}

\section{Generative Adversarial Network}
\label{section:gan}
    As noted earlier, the proposed IoT anomaly detection solution leverages the GAN algorithm to model underlying benign and/or anomalous data traffic patterns. Consider some details on this NN-based scheme first. The GAN concept was originally proposed in \cite{ganog} and uses an adversarial NN-based framework to estimate a generative model. In particular, this setup consists of two NN entities, a generator, $G$, and a discriminator, $D$.  Namely, the former captures the distribution of the data and generates ``fake'' (synthetic) samples.  Meanwhile the latter tries to estimate whether a given sample comes from the actual data, $x$, or the from the latent distribution, $z$. Using these two networks, the GAN approach basically implements a zero-sum game, where the objective of the generator is to produce samples to ``fool'' the discriminator into thinking they came from real data (rather than the model distribution). Overall, studies have shown that this approach is very well-suited for anomaly detection, see \cite{guide},\cite{egan}.  For example, the authors in \cite{guide} use a GAN to detect anomalies in medical image data and accurately predict the early onset of certain diseases. Similarly, \cite{egan} uses a GAN to identify anomalies in non-image data, and tests with the KDD99 and the MNIST datasets also show very promising results.
 
    Since GANs can excel at modeling complex distributions, the proposed  anomaly detection framework herein leverages them to generate normal (benign) and malicious traffic profiles for IoT devices. Specifically, the working hypothesis here is that if a GAN can successfully capture the distribution of given type of data stream, then it should also be able to flag any deviating/outlier behaviors. Hence akin to \cite{guide}, the following optimization problem is considered:
%
%
\begin{multline}
V ( D , E , G ) = 
\mathbb { E } _ { x \sim p _ { X } } \left[ \mathbb { E } _ { z \sim p _ { E } ( \cdot | x ) } [ \log D ( x , z ) ] \right] + \\ 
\mathbb { E } _ { z \sim p _ { Z } } \left[ \mathbb { E } _ { x \sim p _ { G } ( \cdot | z ) } [ 1 - \log D ( x , z ) ] \right]
\end{multline}
    \noindent where $\mathbb { E } _ { x \sim p _ { X } }$ is the distribution of the real dataset (pertaining to benign or malicious IoT samples), $\mathbb { E } _ { z \sim p _ { E } ( \cdot | x ) }$ is the distribution of the latent space captured by \textit{G}, $D(x,z)$ is the discriminator function, $p_{E}$ is the latent distribution, and $p_{X}$ is the actual data distribution.  Meanwhile, $E$ is the encoder that maps the data to the latent space and learns simultaneously along with the generator and discriminator. Hence for anomaly detection using benign training data from the IoT testbed, all malicious samples deviating from the normal traffic distribution will be flagged as anomalous. Conversely, for anomaly detection using malicious training data from the darknet, all normal samples deviating from the malicious traffic distribution will be flagged as anomalous. Note that the similarity between a test sample and the generated latent representation will depend upon how similar it is to the data being used to train the GAN. 

   Finally, akin to the work in \cite{guide}, the GAN algorithm also uses a convex multi-modal loss function comprised of weighted generator and discriminator losses. Namely, the generator loss, $L_G(x)$, measures the similarity between a new sample and the data generated from the latent space, $Z$. Meanwhile, the discriminator loss, $L_D(x)$, measures the fidelity of the generated synthetic samples and is also used to drive the generator to produce samples belonging to the real data distribution. Hence the weighted loss function is given by:\\
\begin{equation}
L ( x ) = \alpha L _ { G } ( x ) + ( 1 - \alpha ) L _ { D } ( x )
\end{equation}
    \noindent where $0 \leq \alpha \leq 1$. Note that the discriminator loss, $L_{D}(x)$, is based on a feature matching methodology that measures the similarity between features of a given sample and that of the generated synthetic data.
 
\section{Empirical Evaluation}
\label{section:results}
    The GAN-based model is now evaluated for anomaly detection in the IoT testbed. First, consider dataset generation and feature selection for ML training. Here the {\tt pfsense} toolkit is used to capture raw packet data in the testbed during ``normal'' operation. This benign traffic is then used to train a GAN. In particular, feature selection is done for bi-directional flows over 3 windows spanning the most recent 50, 100, 500, and 2,000 packets. The mean and standard deviation values for several parameters are then computed for each of these windows, i.e., including the number of packets, packet lengths, packet inter-arrival times, etc. Hence 13 features are extracted for each of the 4 windows, yielding a total of 52 features indexed by the source IP addresses of the IoT devices (see Table \ref{tsig3}). Once the GAN model has been trained using the benign data features, further data collection is also done for attack traffic. Namely, several types of cyberattacks (detailed in Section \ref{attack}) are launched against the IoT devices, and above data collection/processing steps are repeated to extract the relevant features for attack traffic.
   
    As noted earlier, this study also leverages empirical darknet data to further train and validate the GAN approach (i.e., in addition to the GAN trained using data from the IoT testbed). Specifically, the {\tt CAIDA} repository (www.caida.org) is used to extract passive darknet measurements from the network telescope at the University of California San Diego (UCSD). Studies have shown that this data can provide key insights and evidence of Internet-scale unsolicited IoT device behaviors \cite{farooq1}. Hence the working assumption here is that a GAN trained using this data will be able to identify anomalous IoT traffic as belonging to the same distribution as malicious IoT data from the darknet. By extension any normal traffic (generated by clean IoT devices) will appear as ``anomalous'' to this GAN. Accordingly, the benign and malicious IoT traffic datasets are used to test the efficiency of the darknet GAN model as well. In particular, a total of 50,000 benign instances and 3,000 malicious instances of data are generated for each of the three IoT testbed devices.
    
    All ML evaluation is done using the {\tt TensorFlow} model library. Namely, the GAN model is first trained using a mix of testbed and {\tt CAIDA} data (as detailed above). Here, the generator network is formed with 3 dense layers of size 64, 128, and 56 nodes and a \textit{rectified linear unit} (RELU) non-linearity. Similarly, the encoder network has 3 dense layers comprising of 128, 64, and 32 nodes, respectively. Finally, the discriminator network consists of a single dense layer with 128 nodes. Akin to \cite{egan}, two different discriminator models are also used here, i.e., one which attempts to correctly identify real samples and another which attempts to correctly identify latent samples. Furthermore, the batch size is set to 50 samples, and training is done over 100 epochs with a learning rate of 0.1 (10\%). Furthermore, a total of 3,000 instances are used for the test dataset (both benign and malicious samples).
    
\begin{figure}[h]
	\centering
	\includegraphics[width=3.1in, height=2.1in]{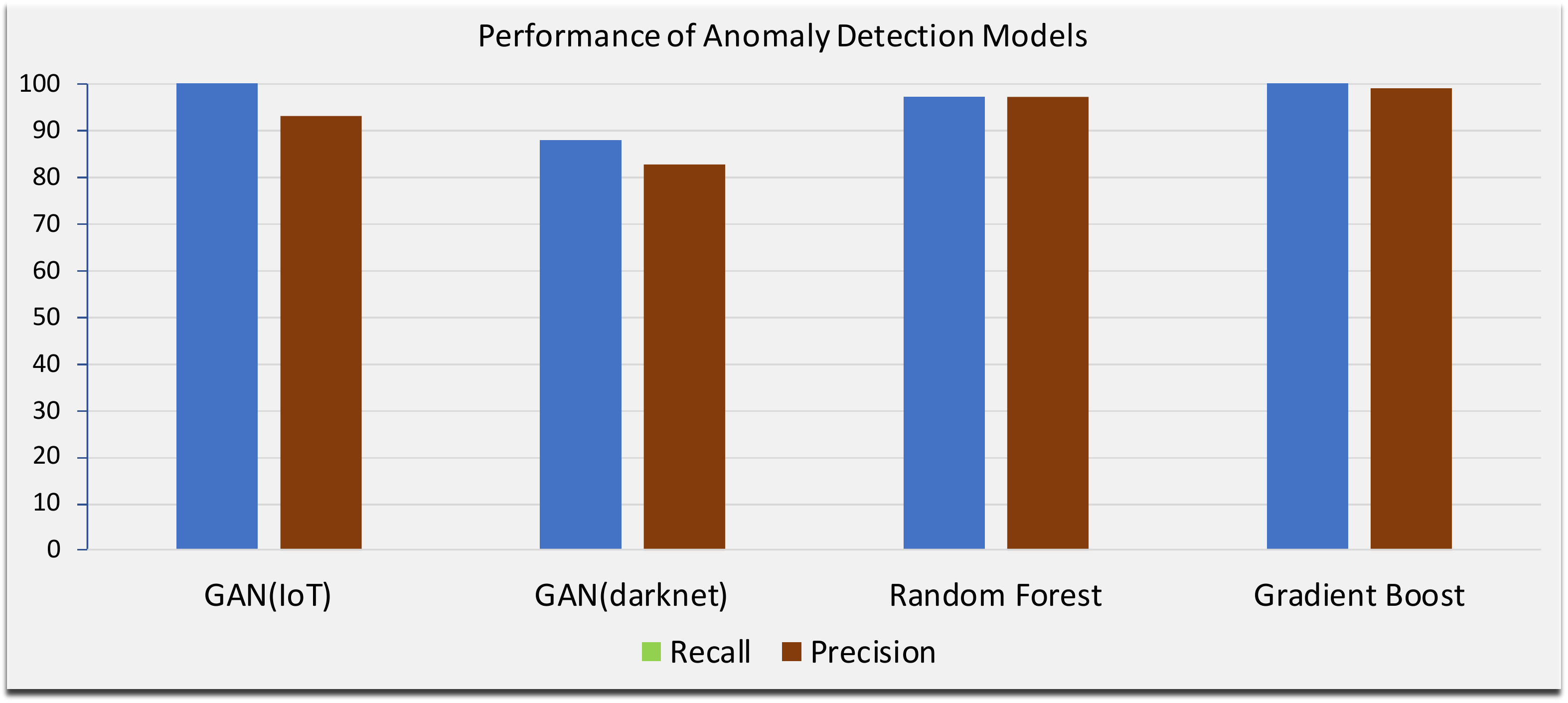}
	\caption{Precision and recall rates} 
	\label{fig:one}
	\end{figure}

    Results for the overall precision and recall performance of the GAN models are presented in Figure \ref{fig:one}. These findings show a precision rate of 100\% and a recall rate of 93\% when the GAN is trained using benign IoT samples from the testbed.  In other words, the latter value implies that no benign IoT samples are mis-identified as malicious. However, approximately 7\% of the malicious samples are still incorrectly classified as benign. Meanwhile, when the GAN model is trained using the {\tt CAIDA} darknet data, it yields slightly lower performance with a precision rate of 97.13\% and a recall of 92\%.  Nevertheless, this performance is still a very impressive considering the fact that the darknet data is largely unrelated to the that of the IoT devices operating in the testbed.

    For comparison purposes, two other classical (non-NN) supervised ML algorithms are also tested, i.e., random forest and gradient boosting. Namely, results for the random forest classifier (Figure \ref{fig:one}) indicate a precision rate of 97.29\% and a recall rate of 97.3\%.  Meanwhile, the gradient boosting classifier gives slightly better performance, with a precision rate of 100\% and a recall rate of 98.90\%. Although these algorithms closely match (even slightly exceed) the performances of the GAN-based schemes, a major drawback here is that they require \textit{both} benign and malicious data samples for training purposes.  By contrast, the GAN models are much more efficient and can be trained using only benign (testbed) or malicious (darknet) samples. Therefore in practical IoT realms, the GAN-based approach will be much more feasible versus these supervised learning methods, i.e., as it only requires data packets from devices directly connected to the network.

\begin{figure}[h]
	\centering
	\includegraphics[width=3.1in, height=2.1in]{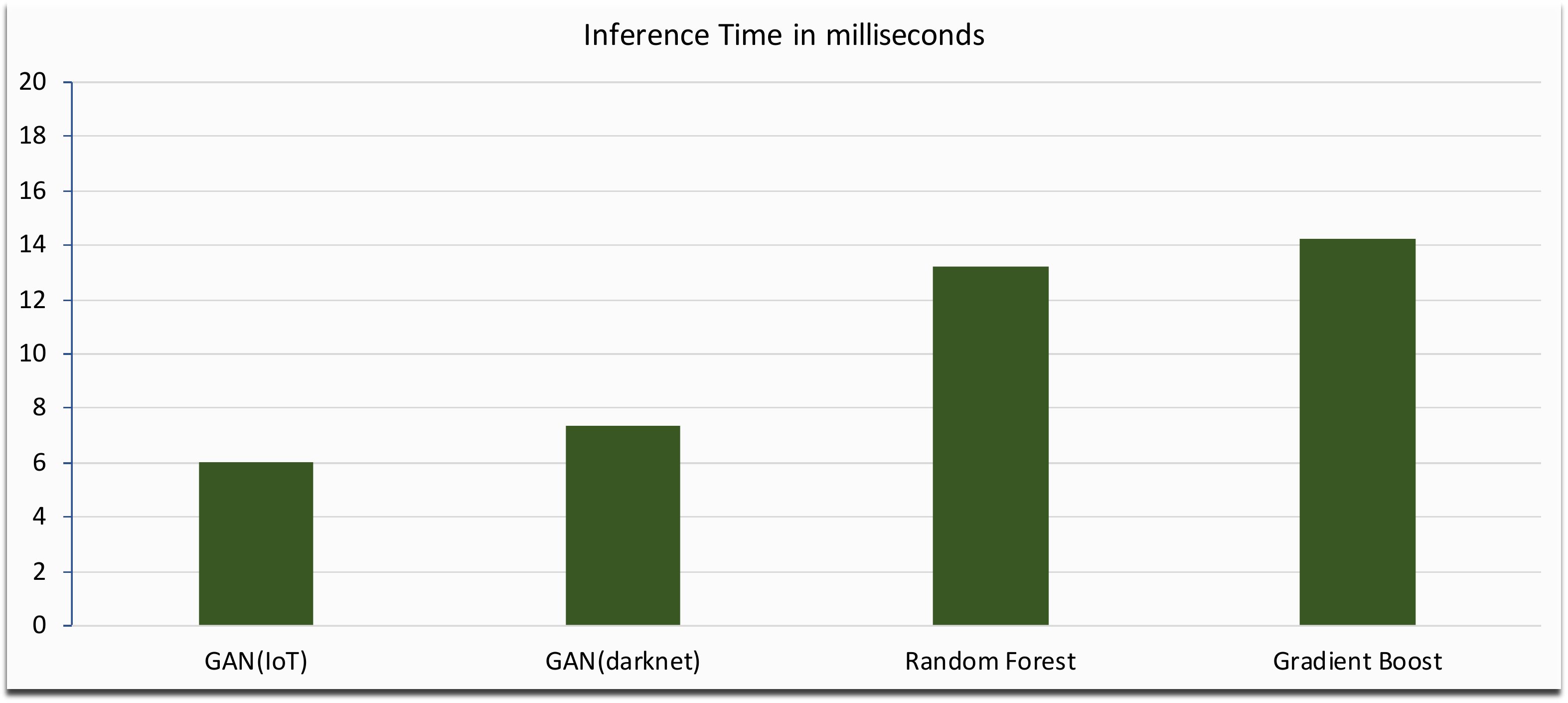}
	\caption{Inference times for test data (ms)} 
	\label{fig:two}
	\end{figure}

    Finally, the inference times of the various ML schemes are also measured and shown in Figure \ref{fig:two}. These findings confirm that the GAN-based algorithms yield the fastest (lowest) times, averaging about 6.05 ms.  Conversely, the random forest and gradient boosting algorithms are notably slower, with average inference times of 12.46 and 14.25 ms, respectively. Again, these findings confirm more efficient run-time operation in practical real-world settings with the GAN-based solution.
	

\section{Conclusions and Future Directions}
\label{section:conclusion}
    \textit{Internet of things} (IoT) paradigms are seeing wide scale traction across many market sectors.  However, the prevalence of billions of IoT devices with limited security provisions presents a very large attack surface for malicious attackers to exploit. As a result, IoT security will remain a major concern for the foreseeable future. Along these lines, this effort presents a novel \textit{generative adversarial network} (GAN) solution to identify threats to IoT devices from both inside and outside a network.  Specifically, this anomaly detection scheme leverages the excellent data mapping capabilities of this algorithm to generate traffic profiles for IoT devices and identify outlier behaviors. In particular, the GAN-based models are trained using both benign IoT traffic data and darknet data from a global network telescope repository. A detailed testbed is also built to implement and validate the proposed scheme using real-world IoT devices and well-known IoT malware threats ({\tt Mirai} and {\tt BashLite}). Overall findings show very promising results with the GAN-based solutions, i.e., in terms of overall precision and recall rates and inference times. Although, several classical supervised learning algorithms give marginally better results in some cases, they are much more compute and data intensive, i.e., requiring both benign and malicious samples for practical training purposes.

    Overall, the contributions of this study can be extended along several key directions.  Foremost, a wider range of physical IoT systems can be incorporated into the testbed.  Although the two types of devices used in this study (Wifi routers and webcams) represent a significant portion of the systems targeted by IoT malware, more specialized and advanced devices can also be considered, e.g., such as industrial \textit{supervisory control and data acquisition} (SCADA) systems. Additionally, improved feature selection can also be done to further improve GAN accuracy and better capture the intricate interactions of advanced exploits. Finally, threat mitigation schemes can also be developed and tested using {\tt pfsense} and other open-source software networking tools.

\bibliographystyle{IEEEtran}
\bibliography{my_bib.bib}

\begin{thebibliography}{10}
\providecommand{\url}[1]{#1}
\csname url@samestyle\endcsname
\providecommand{\newblock}{\relax}
\providecommand{\bibinfo}[2]{#2}
\providecommand{\BIBentrySTDinterwordspacing}{\spaceskip=0pt\relax}
\providecommand{\BIBentryALTinterwordstretchfactor}{4}
\providecommand{\BIBentryALTinterwordspacing}{\spaceskip=\fontdimen2\font plus
\BIBentryALTinterwordstretchfactor\fontdimen3\font minus
  \fontdimen4\font\relax}
\providecommand{\BIBforeignlanguage}[2]{{%
\expandafter\ifx\csname l@#1\endcsname\relax
\typeout{** WARNING: IEEEtran.bst: No hyphenation pattern has been}%
\typeout{** loaded for the language `#1'. Using the pattern for}%
\typeout{** the default language instead.}%
\else
\language=\csname l@#1\endcsname
\fi
#2}}
\providecommand{\BIBdecl}{\relax}
\BIBdecl

\bibitem{netsoft2020}
K.~Friday, E.~Kfoury, E.~Bou-Harb, and J.~Crichigno, ``Towards a unified
  in-network ddos detection and mitigation strategy,'' in \emph{IEEE
  International Conference on Network Softwarization (NetSoft) 2020}, Ghent,
  Belgium, June 2020.

\bibitem{ddosiniot}
K.~Constantinos, K.~Georgios, and S.~Angelos, ``Ddos in the iot: Mirai and
  other botnets,'' \emph{Computer}, vol.~50, pp. 80--84, 2017.

\bibitem{ubootkit}
Y.~Jingyu, G.~Chen, L.~Zaho, L.~Chendong, G.~Jiahua, L.~Guize, and M.~Jinsong,
  ``Ubootkit: A worm attack for the bootloader of iot devices,'' in
  \emph{BlackHat Asia 2018}, Singapore, March 2018.

\bibitem{bruno}
B.~Zarpelao, R.~Miani, C.~Kawakani, and S.~Alvarenga, ``A surevey of intrusion
  detection in the internet of things,'' \emph{Journal of Network and Computer
  Applications}, vol.~84, pp. 25--37, April 2017.

\bibitem{svelte}
S.~Raza, L.~Wallgren, and T.~Voigt, ``Svelte: Real-time intrusion detection in
  the internet of things,'' \emph{Ad Hoc Networks}, vol.~11, pp. 2661--2674,
  November 2013.

\bibitem{deeplearningdistributed}
A.~Diro and N.~Chilamkurti, ``Distributed attack detection scheme using deep
  learning approach for internet of things,'' \emph{Future Generation Computer
  Systems}, vol.~82, pp. 761--768, May 2018.

\bibitem{iotthreatanalysis}
E.~Hodo, X.~Bellekens, A.~Hamilton, P.~Dubouilh, E.~Iorkyase, C.~Tachtatzis,
  and R.~Atkinson, ``Threat analysis of iot networks using artificial neural
  network intrusion detection system,'' in \emph{International Symposium on
  Networks, Computers and Communications (ISNCC 2016)}, Hammamet, Tunisia, May
  2016.

\bibitem{computeintell}
A.~Gupta, O.~Pandey, M.~Shukla, A.~Dadhich, S.~Mathur, and A.~Ingle,
  ``Computational intelligence based intrusion detection systems for wireless
  communication and pervasive computing networks,'' in \emph{2013 International
  Conference on Computational Intelligence and Computing Research}, Madurai,
  India, December 2013.

\bibitem{novelauto}
G.~Tucker~B., ``Novel detection and analysis of deep variational
  autoencoders,'' Ph.D. dissertation, Rochester Institue of Technology, 2018.

\bibitem{farooq1}
F.~Shaikh, E.~Bou-Harb, N.~Neshenko, A.~Wright, and N.~Ghani, ``Internet of
  malicious things: Correlating active and passive measurements for inferring
  and characterizing internet-scale unsolicited iot devices,'' \emph{IEEE
  Communications Magazine}, vol.~56, pp. 170--177, September 2018.

\bibitem{anomalydetectionsurvey}
V.~Chandola, A.~Banerjee, and V.~Kumar, ``Anomaly detection: A survey,''
  \emph{ACM Computing}, vol.~41, pp. 1--58, July 2009.

\bibitem{anoamlydetectionnetwork2}
T.~Lane and C.~Brodley, ``Sequence matching and learning in anomaly detection
  for computer security,'' \emph{AAAI Technical Report WS-97-07}, 1997.

\bibitem{anomalydetectnetwork3}
S.~Sarasama, Q.~Zhu, and J.~Huff, ``Hierarchical kohonenen net for anomaly
  detection in network security,'' \emph{IEEE Transactions on Systems, Man and
  Cybernetics, Part B (Cybernetics)}, vol.~35, pp. 302--312, April 2005.

\bibitem{anomalydetectionwebbased}
C.~Kruegel and G.~Vigna, ``Anomaly detection of web-based attacks,'' in
  \emph{10th ACM Conference on Computer and Communications Security},
  Washington D.C., USA, October 2003.

\bibitem{ultra}
D.~Summerville, K.~Zach, and Y.~Chen, ``Ultra-lightweight deep packet anomaly
  detection for internet of things devices,'' in \emph{IEEE International
  Performance Computing and Communications Conference (IPCCC 2015)}, Nanjing,
  China, December 2015.

\bibitem{iwcms}
F.~Shaikh, E.~Bou-Harb, J.~Crichigno, and N.~Ghani, ``A machine learning model
  for classifying unsoicited iot devices by observing network telescopes,'' in
  \emph{14th International Wireless Communications \& Mobile Computing
  Conference (IWCMC 2018)}, Limassol, Cyprus, June 2018.

\bibitem{netmine}
D.Apiletti, E.Baralis, T.Cerquitelli, and V.~D.Elia, ``Characterizing network
  traffic by means of the netmine framework,'' \emph{Computer Networks},
  vol.~53, pp. 774--789, April 2009.

\bibitem{analysiskdd}
L.~Dhanabal and S.~Shantharajah, ``A study on nsl-kdd dataset for intrusion
  detection system based on classification algorithms,'' \emph{International
  Journal of Advanced Research in Computer and Communication Engineering},
  vol.~6, pp. 446--452, June 2015.

\bibitem{nabiot}
Y.~Median, M.~Bohadana, Y.~Mathov, Y.~Mirsky, A.~Shabtai, D.~Breitenbacher, and
  Y.~Elovici, ``N-baiot: Network-based detection of iot botnet attacks using
  deep autoencoders,'' \emph{IEEE Pervasive Computing}, pp. 12--22,
  July-September 2018.

\bibitem{kali}
G.~Singh and J.~Singh, ``Evaluation of penetration testing tools of kali
  linux,'' \emph{International Journal of Innovations and Advancement in
  Computer Science}, vol.~5, pp. 28--32, September 2016.

\bibitem{pfsense}
D.~Kumar and M.~Gupta, ``Implementation of firewall \& intrusion detection
  system using pfsense to enhance network security,'' \emph{International
  Journal of Electrical Electronics \& Computer Science Engineering}, pp.
  131--137, 2018.

\bibitem{iotevolvepaper1}
N.~Dragoni, A.~Giaretta, and M.~Mazzara, ``The internet of hackable things,''
  in \emph{arXiv:1707.08380}, 2018.

\bibitem{ganog}
I.~Goodfellow, J.~Puget\-Abadie, M.~Mirza, B.~Xu, D.~Warde-Farley, S.~Ozair,
  A.~Courville, and Y.~Bengio, ``Generative adversarial nets,'' in \emph{Neural
  Information Processing and Systems (NIPS 2016)}, Barcelona, Spain, December
  2016.

\bibitem{guide}
T.~Schlegl, P.~Seebock, S.~M.~Waldstein, U.~Schmidt\-Erfurth, and G.~Langs,
  ``Unsupervised anomaly detection with generative adversarial networks to
  guide marker discovery,'' in \emph{Information Processing in Medical Imaging
  2016}, North Carolina, USA, June 2017.

\bibitem{egan}
H.~Zenati, C.~Foo, B.~Lecouat, G.~Manek, and V.~Chandrasekhar, ``Efficient
  gan-based anomaly detection,'' in \emph{arXiv:1802.06222}, 2018.

\end{thebibliography}

\end{document}